\documentclass[12pt]{iopart}
\usepackage{siunitx}
\usepackage{graphicx}
\DeclareSIUnit\Molar{\textsc{m}}
\begin{document}
\bibliographystyle{unsrt}

\title{Flexible graphene transistors for recording cell action potentials }
\author{Benno M. Blaschke$^1$, Martin Lottner$^1$,
Simon Drieschner$^1$, Andrea Bonaccini$^2$, Karolina Stoiber$^1$, Lionel Rousseau$^4$, Ga\"elle Lissourges$^4$ and Jose A. Garrido$^{2,3}$}
\address{\textsuperscript{1}Walter Schottky Institut und Physik-Department, Technische Universit\"at M\"unchen, Am Coulombwall 4, 85748 Garching, Germany}

\address{\textsuperscript{2}ICN2 -- Catalan Institute of Nanoscience and Nanotechnology, Barcelona Institute of Science and Technology and CSIC, Campus UAB, 08193 Bellaterra, Spain}

\address{$^3$ICREA, Instituci\'{o} Catalana de Recerca i Estudis Avan\c{c}ats, 08070 Barcelona, Spain}
\address{\textsuperscript{4}ESIEE-Paris, ESYCOM, University Paris EST, Cit\'{e} Descartes BP99, Noisy-Le-Grand 93160, France}
\ead{joseantonio.garrido@icn.cat}

\begin{abstract}

Graphene solution-gated field-effect transistors (SGFETs) are a promising platform for the recording of cell action potentials due to the intrinsic high signal amplification of graphene transistors. In addition, graphene technology fulfills important key requirements for for in-vivo applications, such as biocompability, mechanical flexibility, as well as ease of high density integration. In this paper we demonstrate the fabrication of flexible arrays of graphene SGFETs on polyimide, a biocompatible polymeric substrate. We investigate the transistor's transconductance and intrinsic electronic noise which are key parameters for the device sensitivity, confirming that the obtained values are comparable to those of rigid graphene SGFETs. Furthermore, we show that the devices do not degrade during repeated bending and the transconductance, governed by the electronic properties of graphene, is unaffected by bending. After cell culture, we demonstrate the recording of cell action potentials from cardiomyocyte-like cells with a high signal-to-noise ratio that is higher or comparable to competing state of the art technologies. Our results highlight the great capabilities of flexible graphene SGFETs in bioelectronics, providing a solid foundation for in-vivo experiments and, eventually, for graphene-based neuroprosthetics. 

\end{abstract}

\section{Introduction}
In recent years, an increasing effort is being dedicated to the development of a new generation of electronic devices that can further advance the interface to living cells and tissue.\cite{Hess.2011b,Xie.2012,Piret.2015,Kim.2010b} Besides improving our understanding of the nervous system and the brain,\cite{Eytan.2006} these devices can be applied in electrically-active prostheses to restore vision,\cite{Zrenner.2011} hearing,\cite{Moore.2009} or to find a solution to damaged motor or sensory functions.\cite{Hochberg.2006}
While some of these applications exclusively rely on the electrical stimulation of cells or tissue, others also require the detection of the electrical activity of the nerve cells. Besides microelectrode array (MEA) technologies\cite{Xie.2012,Buzsaki.2004,Stett.2003,Bruggemann.2011} transistor-based concepts are receiving renewed attention for recording \cite{Fromherz.1991,Dankerl.2009,Hess.2013,Khodagholy.2013,Voelker.2005} due to the advantages they can offer. For instance, their intrinsic signal amplification enabled by the transistor configuration\cite{Hess.2011} and the possibility for downscaling and high density integration in contrast to the MEA technology where the impedance is greatly affected by the electrode size. Furthermore, the development of transistor-based designs could enable a new generation of implants with bidirectional communication capabilities i.e. providing both stimulation and recording, thus allowing an in-situ fine control for electrical stimulation.\cite{Wagenaar.2005} Therefore, there is a need to explore and identify suitable materials for the fabrication of transistors that can be used for recording electrical activity. In this respect the transistor material has to meet several requirements to allow for an efficient and long-lasting interface to living systems: it has to be biocompatible and chemically stable in harsh biological environments, and it has to provide a broad electrochemical potential window to avoid the negative effects of electrochemical reactions.\cite{McCreery.1988} Furthermore, in order to allow for a high sensitivity in the detection of action potentials the material of choice is expected to exhibit good electronic performance, such as high carrier mobility and low intrinsic noise.\cite{Hess.2011b} Materials offering a high capacitance at the electrolyte/transistor interface are also of interest due to the positive influence of the interfacial capacitance on the transistor sensitivity;\cite{Dankerl.2010} additionally, a high capacitance also has a positive effect on the range of gate bias that can be applied to these devices, which is rather limited due to the operation in aqueous electrolytes.\cite{Hess.2013} Lastly, considering the implementation of this technology in real applications, for instance in biomedical implants, it becomes of utmost importance to use materials that allow the fabrication of flexible devices, a requirement needed to lower the mechanical mismatch between the sample and the tissue, thus avoiding the decrease in the device performance due to glial scare formation.\cite{Polikov.2005}
In the past, several materials have been used for cell signal detection in a transistor configuration: silicon,\cite{Fromherz.1991} gallium nitride,\cite{Steinhoff.2005} diamond,\cite{Dankerl.2009} and more recently organic materials\cite{Khodagholy.2013b} and graphene.\cite{Hess.2011b} While the use of materials such as silicon, diamond and gallium nitride introduces enormous technological challenges in terms of device flexibility, organic materials, PEDOT:PSS for instance,\cite{Khodagholy.2013} or novel materials such as graphene\cite{Kim.2009} can be integrated relatively easy into flexible devices. However, many organic materials such as P13\cite{Benfenati.2013} or sexithiophene only provide charge carrier mobilities below \SI{10}{\square\centi\meter\per\volt\per\second}\cite{Buth.2012} and have a relatively high electronic noise. Therefore, high quality chemical vapor deposition (CVD) graphene, offering simultaneously high carrier mobility (well above \SI{e3}{\square\centi\meter\per\volt\per\second}), low electronic noise, high chemical stability and facile integration into flexible devices, appears as a particularly qualified material.\cite{Hess.2013} While the first reports of cell recordings using graphene solution-gated field-effect transistors (SGFETs) based on rigid substrates already demonstrated the great potential of this material,\cite{Hess.2011b} the next challenge is the transfer of that rigid technology to a more suitable flexible one. 
In this paper, we report on the detection of action potential of cardiomyocyte-like HL-1 cells\cite{Claycomb.1998} using flexible graphene based SGFETs. Our work confirms that flexible devices fabricated using CVD graphene can play a significant role in the next generation of implant technologies.
\section{Results and discussion}
The fabrication of the devices, described in detail in the methods section, is carried out on an approximately \SI{10}{\micro\meter} thick polyimide film spin coated on a supporting substrate. In short, metal contacts were evaporated onto the substrate, after which CVD graphene was transferred and the active area of the transistors was defined. Afterwards, a second metal layer was evaporated and the metal lines were covered with an insulating photoresist. In a last step, the device is released from the supporting substrate. The upper panel in figure 1 a) shows a schematic of a released device. The transistors are located in the center and connected to the bond pads via metal feed lines.

\begin{figure}[h]
	\centering
		\includegraphics[width=\textwidth]{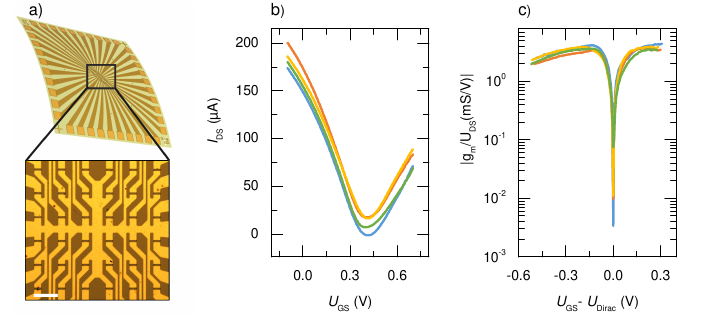}
	
\caption{a) Upper panel: Schematic of a flexible graphene transistor array on a polyimide substrate. Lower panel: Microscope image of 36 transistors of the array with drain and source contacts and the SU8 window. Scale bar is \SI{200}{\micro\meter}. b) Transistor currents of four transistors as a function of the applied gate potential measured in \SI{5}{\milli\Molar} PBS buffer. c) Normalized transconductance for the same transistors ($W$=\SI{20}{\micro\meter}; $L$=\SI{10}{\micro\meter}).}
\label{fig:1}
\end{figure}

A microscope image of a 6x6 transistor array is shown in the lower panel of figure \ref{fig:1} a). The active area of each transistor is \SI{10}{\micro\meter} (length) x \SI{20}{\micro\meter} (width). Firstly, the flexible graphene SGFETs were characterized to compare their performance to existing technologies. The transistor measurements were performed in a \SI{5}{\milli\Molar} phosphate buffered saline (PBS) solution using an Ag/AgCl reference electrode to apply the gate voltage. Figure 1b) shows typical transistor curves in which the drain-source current, $I_{DS}$, was recorded as a function of the gate voltage, $U_{GS}$, while the drain-source voltage was fixed to $U_{DS}$=\SI{100}{\milli\volt}. As expected from the graphene band structure a V-shape curve is observed,\cite{Dankerl.2010} exhibiting the Dirac point (minimum of the curve) around $U_{Dirac}$=\SI{400}{\milli\volt} vs. Ag/AgCl. This indicates p-type doping of the device since for an undoped device a Dirac voltage of about $U_{Dirac}$=\SI{150}{\milli\volt} is expected due to the difference of the work function of graphene (\SI{4.6}{\eV})\cite{Giovannetti.2008} and the Ag/AgCl reference electrode (\SI{4.7}{\eV});\cite{Reiss.1985} the applied $U_{DS}$ should also be considered. Residues from PMMA used during the transfer and interactions with the substrate have been suggested as the origin of the p-type doping of transferred CVD graphene.\cite{Pirkle.2011,Shi.2009}
A key figure of merit of the device performance is the transconductance, $g_m$, which is typically used to quantify the sensitivity of the device and represents the change in the transistor current, $I_{DS}$, induced by a small change in the gate voltage.\cite{Hess.2011} In the particular case of the detection of action potentials with a transistor the electrical activity of a cell in the vicinity of the transistor’s active region will induce a small change of the effective gate voltage, ∆$U_{GS}$, applied to the transistor. Thus, for a given ∆$U_{GS}$, the larger $g_m$, the larger the measured modulation of the transistor current. Figure 1 c) shows the transconductances, normalized by $U_{DS}$, obtained by deriving $I_{DS}$ with respect to $U_{GS}$ in figure \ref{fig:1} b). Values of more than \SI{4}{\milli\siemens\per\volt} are obtained, similar to those of rigid graphene transistors.\cite{Hess.2011b} These values are significantly higher than those reported for transistors based on other technologies, such as silicon, diamond or AlGaN,\cite{Hess.2013} and are comparable to other flexible technologies such as PEDOT:PSS transistors.\cite{Khodagholy.2013} The high transconductance of the graphene SGFETs originates from the combined effect of the interfacial capacitance of the graphene/electrolyte interface, of several \si{\micro\farad\centi\meter^{-2}},\cite{Dankerl.2010} and the high charge carrier mobilities in CVD graphene, of more than \SI{1000}{\square\centi\meter\per\volt\per\second}.\cite{Hess.2011} 

\begin{figure}[h]
	\centering
		\includegraphics[width=\textwidth]{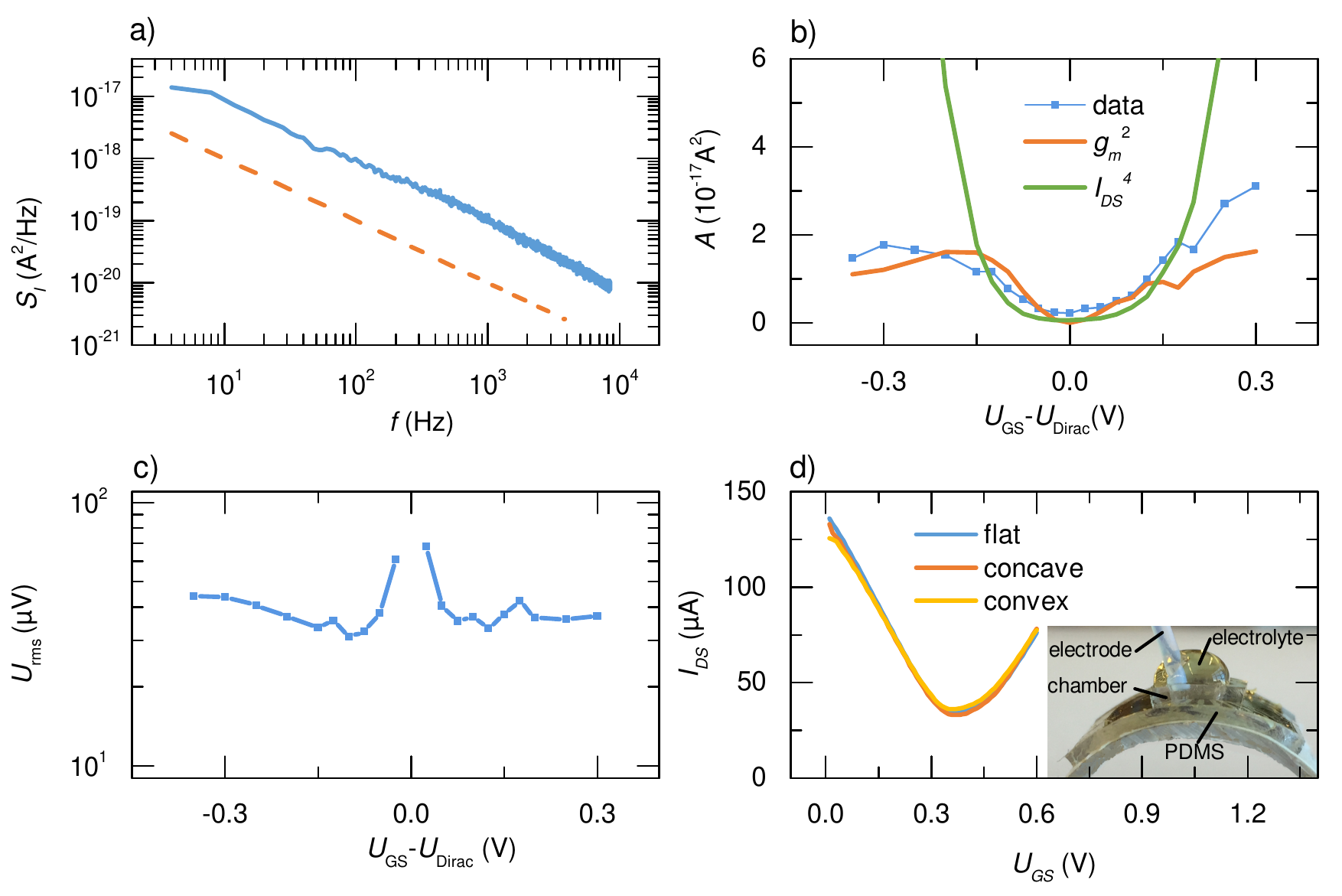}
	\label{fig:2}
\caption{a) Power spectral density of the drain-source current as a function of the frequency (blue). Bias conditions: $U_{GS}$ = \SI{250}{\milli\volt} and $U_{DS}$ = \SI{100}{\milli\volt}.The orange dashed line shows a 1/f dependence. b) Noise parameter A obtained by fitting the power spectral density to a $A/f^b$ law for different $U_{GS}$ values. Measured data points are shown as rectangles. For comparison a ${g_m}^2$ (orange) and a ${I_{DS}}^4$ (green) dependence are shown. c) Calculated effective gate noise of a transistor. d) Drain-source current as a function of the applied gate bias of a device in flat (blue), concave (orange) and convex (yellow) shape. The inset shows a picture of a the device in a convex state.}
\end{figure}

Besides the transconductance, the intrinsic electronic noise of the transistor has to be considered in order to characterize its sensitivity: the noise figure of merit sets the limit for the minimum modulation of the gate, and thus the minimum cell signal that can be detected by the transistor. To assess the noise of the flexible graphene SGFETs, the power spectral density (PSD), $S_I$, of the transistor current was measured in \SI{5}{\milli\Molar} PBS buffer (see methods section for details). Figure \ref{fig:2} a) shows the result of 200 averaged individual spectra obtained for one transistor (bias conditions: $U_{GS}$ =\SI{250}{\milli\volt} and $U_{DS}$ =\SI{100}{\milli\volt}). A 1/f behavior of the power spectral density is observed, as reported previously for rigid graphene SGFETs.\cite{Hess.2011b,Heller.2010} To evaluate the noise performance, the power spectral density is fitted using $S_I =A/{f^b}$, with A and b representing the fitting parameters. Values of $b$ typically range from 0.8 to 1.2. In order to understand the origin of the noise generation mechanism and to identify the most suitable transistor bias conditions in terms of noise, the influence of the gate bias, $U_{GS}$, on the power spectral density has been investigated. Figure \ref{fig:2} b) shows that the noise parameter $A$ as a function of $U_{GS}$ reaches a minimum close to $U_{Dirac}$. For comparison, the graph also shows the $U_{GS}$ dependence of ${g_m}^2$ (orange) and ${I_{DS}}^4$ (green) calculated for the same device. These two dependences have been previously used to discuss the noise mechanisms in graphene transistors.\cite{Heller.2010} On the one hand, a noise parameter A displaying a ${g_m}^2$ dependence has been correlated to a noise mechanism in which charge fluctuations close to the graphene transistor active area are coupled into the device through the interfacial capacitance. On the other hand, a ${I_{DS}}^4$ dependence has been attributed to the noise generated by a serial resistor in the so-called access regions (i.e. the ungated SU8 covered graphene in this transistor design). As figure \ref{fig:2} b) reveals, the noise parameter A mainly follows a ${g_m}^2$ dependence, except for very large $U_{GS}$. Thus, we conclude that the noise is mainly dominated by charge fluctuations close to the active area of the graphene transistor whereas access regions play a minor role. The charge fluctuations are probably related to charge traps in the substrate.\cite{Heller.2010,Bhardwaj.1984} To estimate the sensitivity limit of the devices, i.e. the minimum signal at the gate that can be detected, we calculate the RMS gate noise $U_{RMS}$ using $\left(U_{RMS}\right)^2=\int_{f_1}^{f_2}\frac{S_I}{{g_m}^2}df$ ,\cite{Hauf.2010} with $f_1$=\SI{4}{\hertz} and $f_2$=\SI{3}{\kilo\hertz}, which is the relevant frequency range for biological signals. Figure \ref{fig:2} c) shows the results as a function of $U_{GS}$-$U_{Dirac}$. Values as low as \SI{30}{\micro\volt} are obtained at $U_{GS}$ = \SI{250}{\milli\volt}. At the point of maximum transconductance, $U_{GS}$ =\SI{0.2}{\volt}, it is only slightly higher $U_{rms}$=\SI{33}{\micro\volt}. This is comparable to other device technologies such as silicon and diamond transistors\cite{Hauf.2010} and also comparable to electrode technologies, for instance to graphene or gold electrodes.\cite{Kuzum.2014}

In order to investigate the device performance under bending conditions, a sample was fixed on a thin PDMS sheet to facilitate the handling. The device was bent to concave and convex shape with a bending radius of approx. \SI{10}{\milli\meter} and \SI{12}{\milli\meter} respectively. The drain-source current $I_{DS}$ was measured as a function of the gate voltage $U_{GS}$ in \SI{5}{\milli\Molar} PBS buffer while the drain-source voltage was fixed to $U_{DS}=$\SI{200}{\milli\volt}. Figure \ref{fig:2} d) shows the obtained transistor curves and a transistor curve in flat shape for comparison. The transistor's transconductance and Dirac point show no discernible dependence on the device curvature. The inset of figure \ref{fig:2} d) shows a picture of the device in convex shape.

In order to demonstrate the detection of action potentials with our flexible graphene transistor arrays, experiments were performed with HL-1 cells cultured onto the chip. After plating (see methods section) a confluent layer of HL-1 cells formed on the device. The cell culture did not degrade the transistor performance. 
 \begin{figure}[h]
	\centering
		\includegraphics[width=\textwidth]{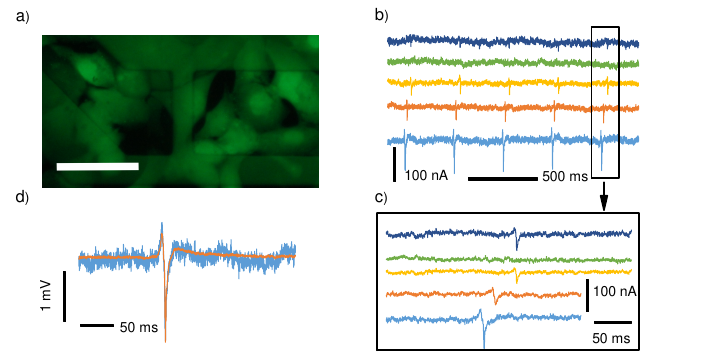}

\caption{a) Fluorescence image of HL-1 cells on a flexible graphene SGFET. Scale bar is \SI{60}{\micro\meter}. b) Current of several graphene transistors showing action potential recordings of HL-1 cells. c) Zoom into a single action potential from b). d) Calculated change in the effective gate potential of a transistor from c) for a single recording (blue) and averaging 47 action potentials (orange).}
	\label{fig:3}
\end{figure}
Figure \ref{fig:3} a) shows a fluorescence image of a device with a cell culture before confluence (cells were stained with Fluo-4 AM, as described in the methods section). The gold feed lines appear black, since in contrast to the polyimide substrate they show no fluorescence. As can be seen in the figure \ref{fig:3} a), a cell lies directly between the source and drain contacts of the transistor. In confluent HL-1 cell layers, spontaneous action potentials triggered by pacemaker cells spread across the layer.\cite{Hofmann.2010} In order to detect these action potentials the transistors were operated close to the point of maximum transconductance (constant $U_{DS}$ = 30 mV and $U_{GS}$ = 200 mV) and the setup recorded the transistor currents (see methods section for details). Figure \ref{fig:3} b) shows the recorded currents of five different transistors in this array. In four of the transistors action potentials were recorded with a frequency of approx. \SI{3}{\hertz} to \SI{4}{\hertz}, in agreement with previously reported beating frequencies of HL-1 cells.\cite{Xie.2012,Fahrenbach.2007} The recording of the transistor exhibiting no action potentials is provided to demonstrate that the recorded action potentials signals are not caused by the recording setup or related to external noise. Figure \ref{fig:3} c) shows that the action potentials are not recorded at the very same time by every transistor, which results from the propagation of the cell signals in the confluent cell layer. It should also be noted that both the action potential amplitude and shape vary from transistor to transistor, which is due to the different coupling between the cells and each transistor.\cite{Fromherz.1991} 
Using the transconductance the recorded variation of the current can be converted into a modulation of the effective gate potential. The signal to noise ratio, defined as signal peak-to-peak amplitude divided by the rms noise, is 19. This is comparable to other state of the art technologies such as nanocavity electrode arrays.\cite{Hofmann.2010} Additionally, the signals of several action potentials can be averaged since consecutive spikes are expected to have the same shape. By means of this averaging procedure the signal-to-noise ratio can be increased, as shown in figure \ref{fig:3} d) where a single action potential (blue) and the average of 47 signals (orange) are depicted.
\section{Conclusion}
In conclusion, we have demonstrated the fabrication of arrays of flexible solution-gated transistors based on CVD graphene on polyimide substrates. The transistors show high transconductance and low electronic noise and do not degrade during bending experiments. After the successful culture of electrogenic HL-1 cells we were able to record action potentials from the cells with excellent signal-to-noise ratio. Future experiments should aim at in-vivo recordings of cell activity to pave the way for the development of a future generation of electrically-active flexible implants.
\section{Methods}
\subsection*{Transistor fabrication}
Polyimide (PI2611, HD Microsystems) was spin coated onto a silicon/silicon dioxide substrate to achieve a \SI{10}{\micro\meter} thick layer. Next, Ti/Au metal leads were fabricated by a lift-off process and evaporation. Afterwards, CVD-grown graphene was transferred from a copper foil to the substrate. After defining the active area of the transistor by graphene etching in an oxygen plasma, the sample was annealed at \SI{570}{\kelvin} in a forming gas atmosphere. Top contacts of Ni/Au were deposited by evaporation and defined by an etching process. After another annealing step, an approx. \SI{2}{\micro\meter} thick SU8 photoresist (GM1040 Gersteltec) layer was spin coated and structured such that only the graphene between source and drain contact is exposed to the electrolyte. In the presented device design, approx. \SI{1}{\micro\meter} long stripes of graphene are covered with SU8 on each side of the window (access region). Afterwards, the device was released from the silicon wafer in DI water. For most of the measurements and the cell culture the device was fixed on a custom made PCB and an electrolyte container was placed on top.
\subsection*{Transistor characterization}
A self-built setup was used for the transistor characterization. Operational amplifiers transformed the transistor currents to voltages and amplified them. A NI DAQCard recorded the voltages using LabVIEW. The gate voltage was applied by a flexible Ag/AgCl reference electrode (World Precision Instruments).
\subsection*{Noise characterization}
The sample was placed in a custom setup in a Faraday cage. A current-voltage converter based on an operational amplifier feedback loop transformed the transistors currents into voltages; it further amplified and low pass filtered them at \SI{16}{\kilo\hertz}. A second amplification stage removed the DC component of the signal and amplified it by a factor of 100. A NI DAQCard acquired the output voltages and a LabVIEW program calculated the PSD.
\subsection*{Cell culture}
HL-1 cells were obtained from Louisiana State University Health Science Center, New Orleans, LA, USA.\cite{Claycomb.1998} The culture was done in Claycomb medium with 10 \% fetal bovine serum, penicillin (\SI{100}{\micro\gram\per\milli\litre}), norepinephrine (\SI{0.1}{\gram\per\mol}) and L-Glutamine (\SI{2}{\gram\per\mol}) in a fibronectin/gelatin coated petri dish, incubated at \SI{310}{\kelvin}. When beating of the cells was observed, they were subcultured or plated on the chips as follows: The cells were detached from the petri dish using trypsin. The enzymatic activity was then stopped by adding trypsin inhibitor. Cells were then centrifuged for two minutes at approx. \SI{120}{\gram}. After resuspension the cells in culture medium, they were plated in a new petri dish or on the chips. Previously, the chips were sterilized in ethanol (70\%) for 20 minutes, coated with fibronectin and carefully cleaned with PBS buffer. The medium was exchanged daily. The culture became confluent within two to four days after subculture. 
For the fluorescence labeling, the devices were first washed two times with PBS buffer. Next, the sample was incubated for 45 minutes in PBS (\SI{350}{\micro\litre}), Fluo\--4 AM (\SI{0.9}{\micro\litre}, Life technologies) and Pluronic F\--127 (\SI{25}{\micro\litre}, Sigma Aldrich). After washing three times with PBS buffer, the imaging was performed in PBS.
\subsection*{Cell measurements} A home-built system was used for the measurement of cell action potentials. The transistors were operated at a constant source-drain and gate voltage. Operational amplifier feedback loops transformed $I_{DS}$ into a voltage and amplified it. The signal was also band-pass filtered between \SI{0.5}{\hertz} and \SI{10}{\kilo\hertz}. A NI DAQCard acquired the voltage at a sampling frequency of $f_s$ = \SI{30}{\kilo\hertz}. Using MATLAB, the data were also band-pass filtered from \SI{4}{\hertz} to \SI{3}{\kilo\hertz} after digitalization.

\section{Acknowledgements}
We acknowledge M. Seifert, L.H. Hess, M. Sachsenhauser, and M. Prexl for fruitful discussions. JAG, BMB, ML and SD acknowledge support by the German Research Foundation (DFG) in the framework of the Priority Program 1459 ‘Graphene’, the Nanosystems Initiative Munich (NIM), the European Union under the NeuroCare FP7 project (Grant Agreement 280433) and the Graphene Flagship (Contract No. 604391)

\end{document}